\documentclass[showpacs,showkeys,preprintnumbers,floatfix,superscriptaddress]{revtex4}

\usepackage{graphicx}
\usepackage{graphics}
\usepackage{amsmath}
\usepackage{amssymb}

\begin{document}

\title{On the self-consistent general relativistic equilibrium equations of neutron stars}

\author{Jorge A. Rueda}
\affiliation{Dipartimento di Fisica and ICRA, Universita di Roma ``La Sapienza'', P.le Aldo Moro 5, I--00185 Rome, Italy}
\affiliation{ICRANet, P.zza della Repubblica 10, I--65122 Pescara, Italy}

\author{Remo Ruffini}
\email{ruffini@icra.it}
\affiliation{Dipartimento di Fisica and ICRA, Universita di Roma ``La Sapienza'', P.le Aldo Moro 5, I--00185 Rome, Italy}
\affiliation{ICRANet, P.zza della Repubblica 10, I--65122 Pescara, Italy}
\affiliation{ICRANet, Universite de Nice Sophia Antipolis, Grand Chateau, BP 2135, 28 Avenue de Valrose, 06103 NICE CEDEX, France}
\author{S.-S. Xue}
\affiliation{ICRANet, P.zza della Repubblica 10, I--65122 Pescara, Italy}

\date{\today}

\begin{abstract}

We address the existence of globally neutral neutron star configurations in contrast with the traditional ones constructed by imposing local neutrality. The equilibrium equations describing this system are the Einstein-Maxwell equations which must be solved self-consistently with the general relativistic Thomas-Fermi equation and $\beta$-equilibrium condition. To illustrate the application of this novel approach we adopt the Baym, Bethe, and Pethick (1971) strong interaction model of the baryonic matter in the core and of the white-dwarf-like material of the crust. We illustrate the crucial role played by the boundary conditions satisfied by the leptonic component  of the matter at the interface between the core and the crust. For every central density an entire new family of equilibrium configurations exists for selected values of the Fermi energy of the electrons at the surface of the core. Each such configuration fulfills global charge neutrality and is characterized by a non-trivial electrodynamical structure. The electric field extends over a thin shell of thickness $\sim \hbar/(m_e c)$ between the core and the crust and becomes largely overcritical in the limit of decreasing values of the crust mass.

\end{abstract}



\maketitle

One of the fundamental issues in physics and astrophysics is the creation of an electron-positron plasma in overcritical electric fields larger than $E_c = m^2_e c^3/(e \hbar)$ (see \cite{physrep} and references therein). Basic progress toward the understanding of the thermalization process of such a plasma have been achieved \cite{veresh}. The existence of such an electron-positron plasma has a central role in a variety of problems ranging from the acceleration process in gamma ray bursts (GRBs) \cite{physrep} to the sharp trigger process in supernova phenomena \cite{vissani1,vissani2}. This has motivated us to reconsider the standard treatment of neutron stars in order to find a theoretical explanation for the emergence of a wide variety of astrophysical situations involving such overcritical electric fields. In a classic article Baym, Bethe and Pethick \cite{BBP} presented the problem of  matching to the crust in a neutron star a liquid  core composed of $N_n$ neutrons, $N_p$ protons and $N_e$ electrons. After discussing various aspects of the problem  they conclude: `the details of this picture requires further elaboration; this is a situation for which the Thomas-Fermi method is useful.' In this letter we focus on relaxing the traditional condition of local charge neutrality $n_e=n_p$, which appears to have been assumed only for mathematical convenience without any physical justification. Instead, we adopt the more general condition of global charge neutrality $N_e=N_p$. The corresponding equilibrium equations then follow from self-consistent solution of the relativistic Thomas-Fermi equation, the Einstein-Maxwell equations and the $\beta$-equilibrium condition, properly expressed in general relativity.

The pressure and the density of the core are mainly due to the baryons while the pressure of the crust is mainly due to the electrons with the density due to the nuclei and possibly some free neutrons due to neutron drip (see e.g. \cite{BBP}). The boundary conditions determined by the matching of the electron distribution in the core with that of the electrons of the crust are fundamental for the self-consistent construction of the equilibrium configurations.

We consider the case of a non-rotating neutron star with metric
\begin{equation}
ds^2 = e^\nu dt^2 - e^\lambda dr^2 - r^2d\theta^2 - r^2 \sin^2\theta d\phi^2\, ,
\end{equation}
where $\nu$ and $\lambda$ are functions only of $r$. We assume units where $G=\hbar=c=1$ and let $\alpha$ denote the fine structure constant. As usual we define the mass of the star $M(r)$  by $e^{-\lambda} = 1-2 M/r + r^2 E^2$, and denote the Coulomb potential by $V(r)$, which determines the electric field $E=e^{-(\nu+\lambda)/2}V'$, where a prime indicates the radial derivative. The combined energy-momentum tensor of the matter and fields $T_{\mu \nu}$ is given by
\begin{equation}\label{eq:Tab}
T^{\alpha}_{\beta}={\rm diag}({\cal E}+{\cal E}^{em},-P-P^{em},-P+P^{em},-P+P^{em})\, ,
\end{equation}
where ${\cal E}^{\rm em}=-P^{em}=E^2/(8\pi)$, and ${\cal E}$ and $P$ are the energy density and pressure of matter. With all the above definitions, the time-independent Einstein-Maxwell field equations read
\begin{align}
&M'= 4 \pi r^2 {\cal E} + 4 \pi e^{\lambda/2} r^3 e E (n_p - n_e)\, ,\label{eq:Gab1}\\
&e^{-\lambda}\left( \frac{\nu'}{r}+\frac{1}{r^2} \right) - \frac{1}{r^2} = - 8 \pi \,T^1_1 \, ,\label{eq:Gab2}\\
&e^{-\lambda} \left[ \nu'' + (\nu'-\lambda')\left( \frac{\nu'}{2}+\frac{1}{r} \right) \right] = - 16 \pi T^2_2\, , \label{eq:Gab3}\\
&(e V)'' + (e V)' \left[ \frac{2}{r} - \frac{(\nu' + \lambda')}{2}
\right] = - 4 \pi \alpha e^{\nu/2} e^{\lambda} (n_p - n_e)\, .\label{eq:Gab4}
\end{align}

In order to close the system of equilibrium equations, the condition of local charge neutrality $n_e=n_p$ has been traditionally imposed for mathematical simplicity. In this case the problem is reduced to solving only the Einstein equations for a Schwarzschild metric. When this condition is relaxed, imposing only global charge neutrality $N_e=N_p$, we need to satisfy the Einstein-Maxwell equations (\ref{eq:Gab1})--(\ref{eq:Gab4}). In order to impose global charge neutrality as well as quantum statistics on the leptonic component, the general relativistic Thomas-Fermi equation must also be satisfied.

The general relativistic electron Fermi energy is given by
\begin{equation}\label{eq:electroneq}
E^F_e = e^{\nu/2} \mu_e - e V = \rm{constant}\, ,
\end{equation}
where $\mu_e=\sqrt{(P^F_e)^2+m^2_e}$  and $P^F_e=(3 \pi^2 n_e)^{1/3}$ are respectively the chemical potential and Fermi momentum of degenerate electrons. From Eqs. (\ref{eq:Gab4}) and (\ref{eq:electroneq}) we obtain the general relativistic Thomas-Fermi equation
\begin{equation}\label{eq:relTF}
(e V)'' + (e V)' \left[ \frac{2}{r} - \frac{(\nu' + \lambda')}{2}
\right] = - 4 \pi \alpha e^{\nu/2} e^{\lambda} \left\{ n_p - \frac{e^{-3 \nu/2}}{3 \pi^2}[(E^F_e + e V)^2 - m^2_e e^{\nu}]^{3/2}\right\}\, .
\end{equation}
The $\beta$-equilibrium condition is expressed by
\begin{equation}\label{eq:betaeq}
\mu_n = \mu_e + \mu_p\, .
\end{equation}

In order to take into account the effect of the compression of the crust on the leptonic component of the core we solve the equilibrium conditions for the core within a Wigner-Seitz cell \cite{letter2}. The radius 
$R_{WS}$ of this cell  determines the Fermi energy of the electrons of the core which has to be matched with the Fermi energy of the leptonic component of the crust. Global charge neutrality is specified by
\begin{equation}\label{eq:neutrality}
\int_{0}^{R_{WS}} e^{\lambda/2} n_p d^3r  = \int_{0}^{R_{WS}} e^{\lambda/2} n_e d^3r \, .
\end{equation}
From Eqs.~(\ref{eq:betaeq}) and (\ref{eq:neutrality}) we can determine self-consistently the proton, neutron, electron fractions inside the core as well as the radius $R_{WS}$ of the Wigner-Seitz cell of the core \cite{letter2}.

The coupled system of equations consisting of the Einstein-Maxwell equations (\ref{eq:Gab1})--(\ref{eq:Gab3}), the general relativistic Thomas-Fermi equation (\ref{eq:relTF}), the $\beta$-equilibrium condition (\ref{eq:betaeq}) along with the constraint  (\ref{eq:neutrality}) needs, in order to be closed, an equation of state (EOS) for the baryonic component in the core and for the leptonic component of the crust.

In order to illustrate the application of this approach we adopt, as an example, the Baym, Bethe, and Pethick (BBP) \cite{BBP} strong interaction model for the baryonic matter in the core as well as for the white-dwarf-like material of the crust. The general conclusions we reach will in fact be independent of the details of this model.

At the neutron star radius $r=R$, all the electrodynamical quantities must be zero as a consequence of the global charge neutrality condition. Consequently, we have a matching condition with the Schwarzschild spacetime which imposes the boundary condition
\begin{equation}\label{eq:matching}
e^{\nu(R)/2} = \sqrt{1- \frac{2 M(R)}{R}}\, .
\end{equation}

The boundary conditions at the center correspond to $M(0)=0$ and the regularity condition to $n_e(0)=n_p(0)$. From the $\beta$-equilibrium condition (\ref{eq:betaeq}), we can evaluate the central chemical potentials $\mu_e(0)$, $\mu_p(0)$, and $\mu_n(0)$, or equivalently, the central number densities $n_e(0)$, $n_p(0)$, and $n_n(0)$ \cite{jorgePRD}. From Eq.~(\ref{eq:electroneq}) we also have the relation
\begin{equation}\label{eq:centralgravity}
e^{\nu(0)/2} = \frac{E^F_e + e V(0)}{\mu_e(0)}\, .
\end{equation}

Having determined the boundary conditions at infinity and at the center, we turn now to the matching conditions at the surface of the core. Following  BBP \cite{BBP}, the neutron profile at the core-crust interface is given by
\begin{equation}\label{eq:neutronsurface}
n_n(z) = n^{crust}_n + (n^{core}_n - n^{crust}_n) f(z/b)\, .
\end{equation}
We have defined $n^{core}_n = n_n(R_c)$ and $n^{crust}_n = n_n(R_{WS})$. Here $R_c$ is the radius of the core defined as the point where the rest-mass density reaches the nuclear saturation density, i.e., $\rho(R_c) = \rho_0 \simeq 2.7 \times 10^{14}$ g cm$^{-3}$ \cite{BBP}. The function $f(z/b)$ satisfies $f(-\infty) =1$, $f(\infty)=0$, where $b \simeq (n^{core}_n - n^{crust}_n)^{-1/3} \simeq 1/m_\pi$ \cite{BBP}. As proposed by BBP, an appropriate choice for the function $f(z/b)$ is the Woods-Saxon profile $f(z/b) = (1+e^{z/b})^{-1}$. The $z$-coordinate lines are perpendicular to the sharp surface separating two semi-infinite regions (core and crust) in the planar approximation \cite{BBP}; the neutron density approaches $n^{core}_n$ as $z\to -\infty$ and $n^{crust}_n$ as $z\to \infty$.

The matching between the core and the crust occurs at the radius $R_{WS}$, where we have $V'(R_{WS}) = 0$ by virtue of the global neutrality condition given by Eq.~(\ref{eq:neutrality}), and we also choose the value of the Coulomb potential $V (R_{WS}) = 0$. From the electron chemical potential $\mu_e (R_{WS})$ at the edge of the crust, we calculate the corresponding neutron chemical potential $\mu_n(R_{WS})$ according to the BBP treatment. If $\mu_n(R_{WS}) - m_n > 0$, neutron drip occurs. In this case, the pressure is due to the neutrons as well as to the leptonic component, so we have the inner crust (see Table \ref{table:physicalvalues} and \cite{BBP,jorgePRD} for details). For larger values of the radii, i.e., for $r > R_{WS}$ the condition $\mu_n(r) - m_n < 0$ is reached at $\rho_{drip} \simeq 4.3 \times 10^{11}$ g cm$^{-3}$ and there the outer crust starts, with the pressure only determined by the leptonic component. If $\mu_n(R_{WS}) - m_n < 0$, only the outer crust exists.

For a fixed central rest-mass density $\rho(0) \simeq 9.8 \times 10^{14}$ g cm$^{-3}$ and selected values of $E^F_e$ we have integrated the system of equations composed by the general relativistic Thomas-Fermi equation (\ref{eq:relTF}), the $\beta$-equilibrium condition (\ref{eq:betaeq}), the Einstein-Maxwell equations (\ref{eq:Gab1})-(\ref{eq:Gab3}), with the constraint of overall neutrality (\ref{eq:neutrality}).

We found that although the electrodynamical properties of the core are very sensitive to the Fermi energy of the electrons (see Table \ref{table:physicalvalues} for details), the bulk properties of the core like its mass and radius are not sensitive to the value of $E^F_e$. This is perfectly in line with the results of Ruffini et al.\ in \cite{letter2}.

\begin{table}[h]
\begin{equation*}
\begin{array}{|c|c|c|c|c|c|c|c|c|c|}
\hline 
\frac{E^F_e}{m_\pi} & \frac{M(R_c)}{M_\odot} & R_c ({\rm km}) & \frac{e V(0)}{m_\pi} & \frac{e V(R_c)}{m_\pi} & \frac{E_{max}}{E_c} & \frac{\rho_{crust}}{\rho_{drip}} & \frac{M_{crust}}{M_\odot} & \Delta^{ic}_r ({\rm m}) & \Delta^{oc}_r ({\rm km})\\
\hline
0.10 & 0.24 & 5.98 & 1.085 & 0.60 & 388.72 & 0.125 & 2.45\times 10^{-6} & 0.00 & 0.797 \\
\hline
0.15 & 0.24 & 5.98 & 0.985 & 0.55 & 381.04 & 0.384 & 1.15\times 10^{-5} & 0.00 & 1.251 \\
\hline
0.20 & 0.24 & 5.98 & 0.935 & 0.50 & 370.89 & 1.000 & 4.45\times 10^{-5} & 0.00 & 1.899 \\
\hline
0.30 & 0.24 & 5.98 & 0.835 & 0.40 & 346.67 & 46.19 & 4.83\times 10^{-5} & 1.89 & 1.899 \\
\hline
0.35 & 0.24 & 5.98 & 0.785 & 0.35 & 332.43 & 80.83 & 5.42\times 10^{-5}& 2.85 & 1.899 \\
\hline
\end{array}
\end{equation*}
\caption{Results of the numerical integration of the BBP model for selected values of $E^F_e$ for $\rho(0) \simeq 9.8 \times 10^{14}$ g cm$^{-3}$. We show the mass and radius of the core $M(R_c)$ and $R_c$, the Coulomb potential at the center and at the core surface $eV(0)$ and $eV(R_c)$, the peak of the electric field in the core-crust interface $E_{max}$, the rest-mass density at the edge of the crust $\rho_{crust} \equiv \rho(R_{WS})$, the mass of the crust $M_{crust}$, and the inner and outer crust thickness $\Delta^{ic}_r$ and $\Delta^{oc}_r$.}\label{table:physicalvalues}
\end{table}

\begin{figure}[h]
\centering
$$
\begin{array}{cc}
\includegraphics[scale=0.45]{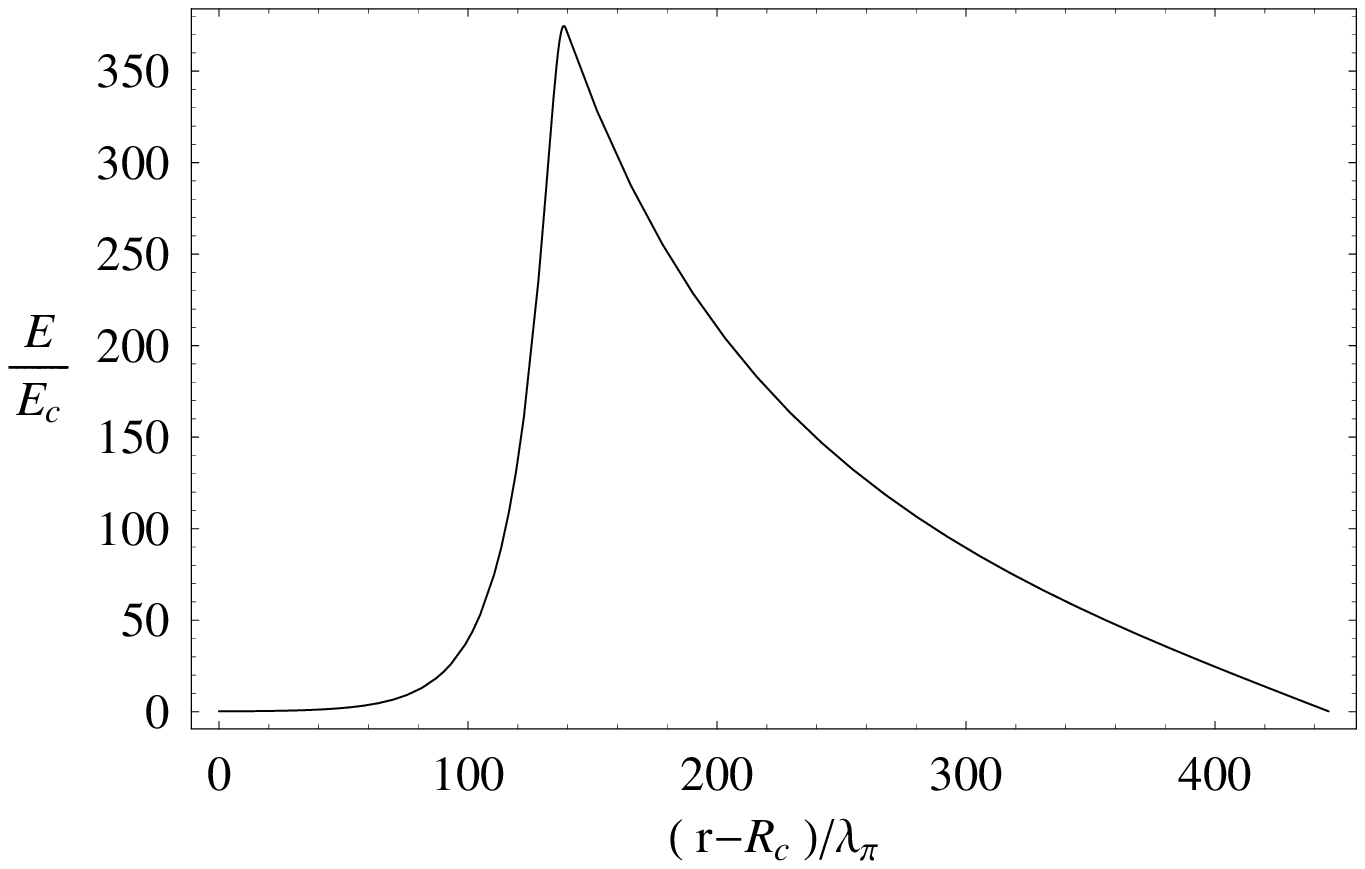} & \includegraphics[scale=0.45]{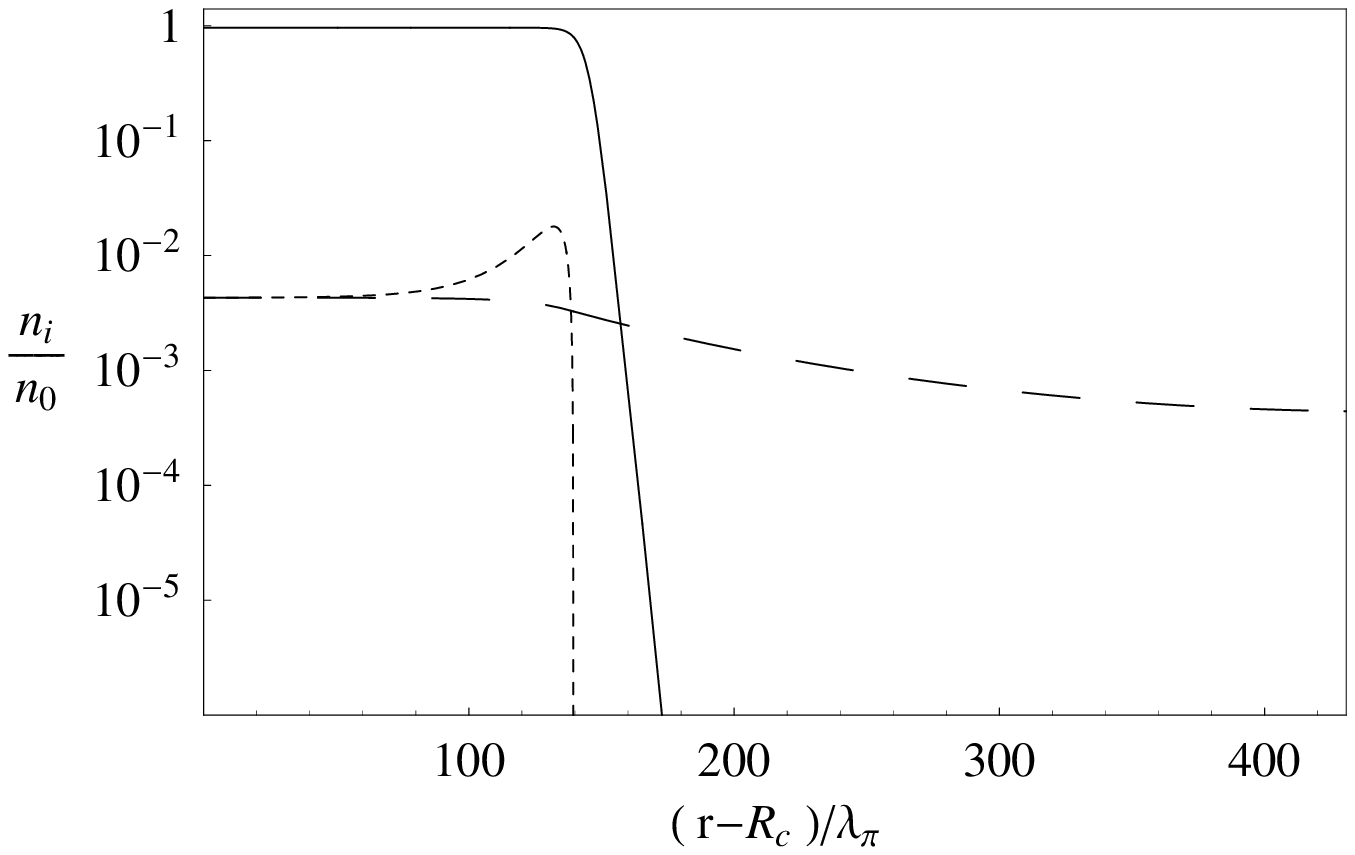} \\
\includegraphics[scale=0.45]{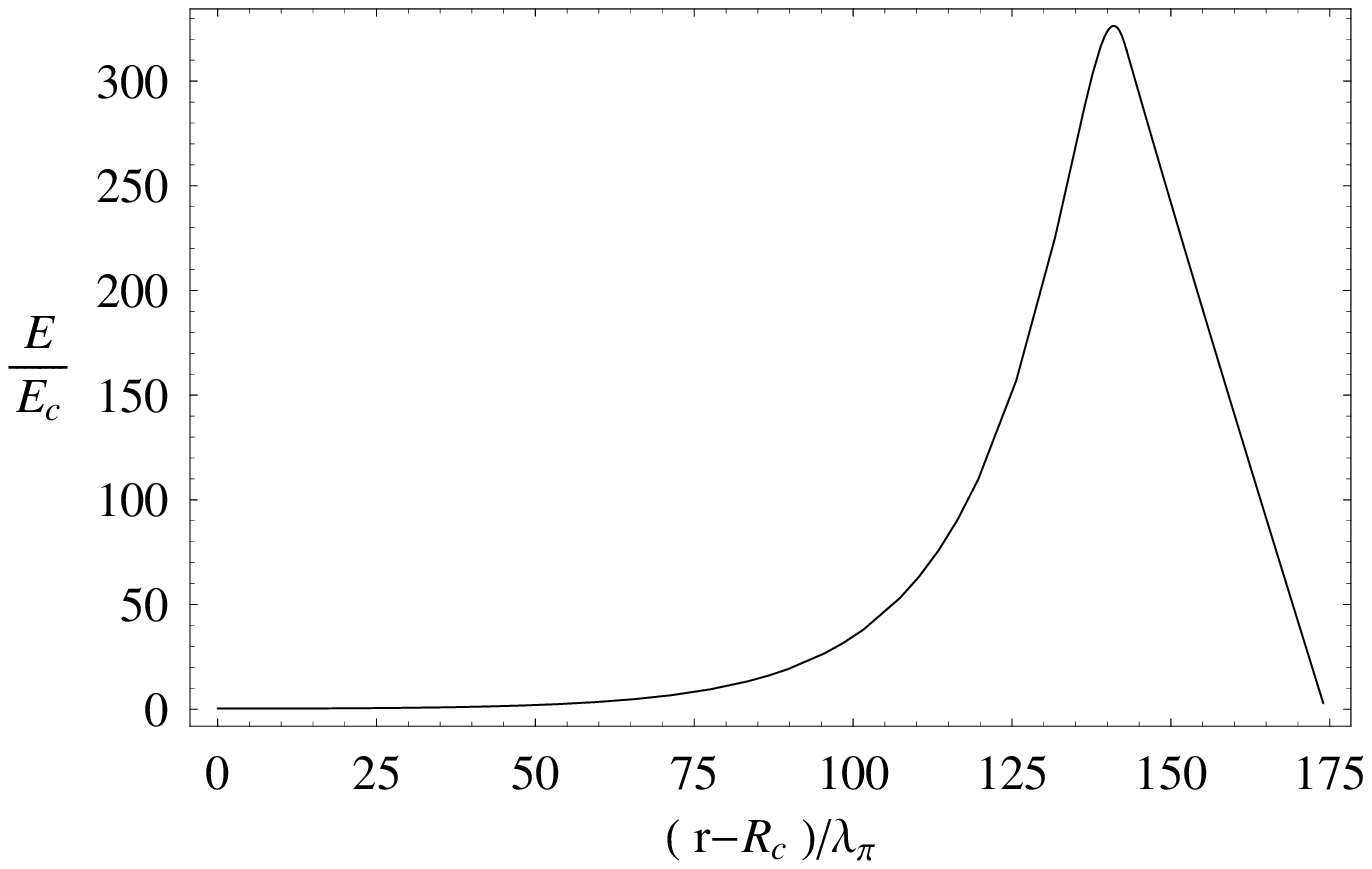} & \includegraphics[scale=0.45]{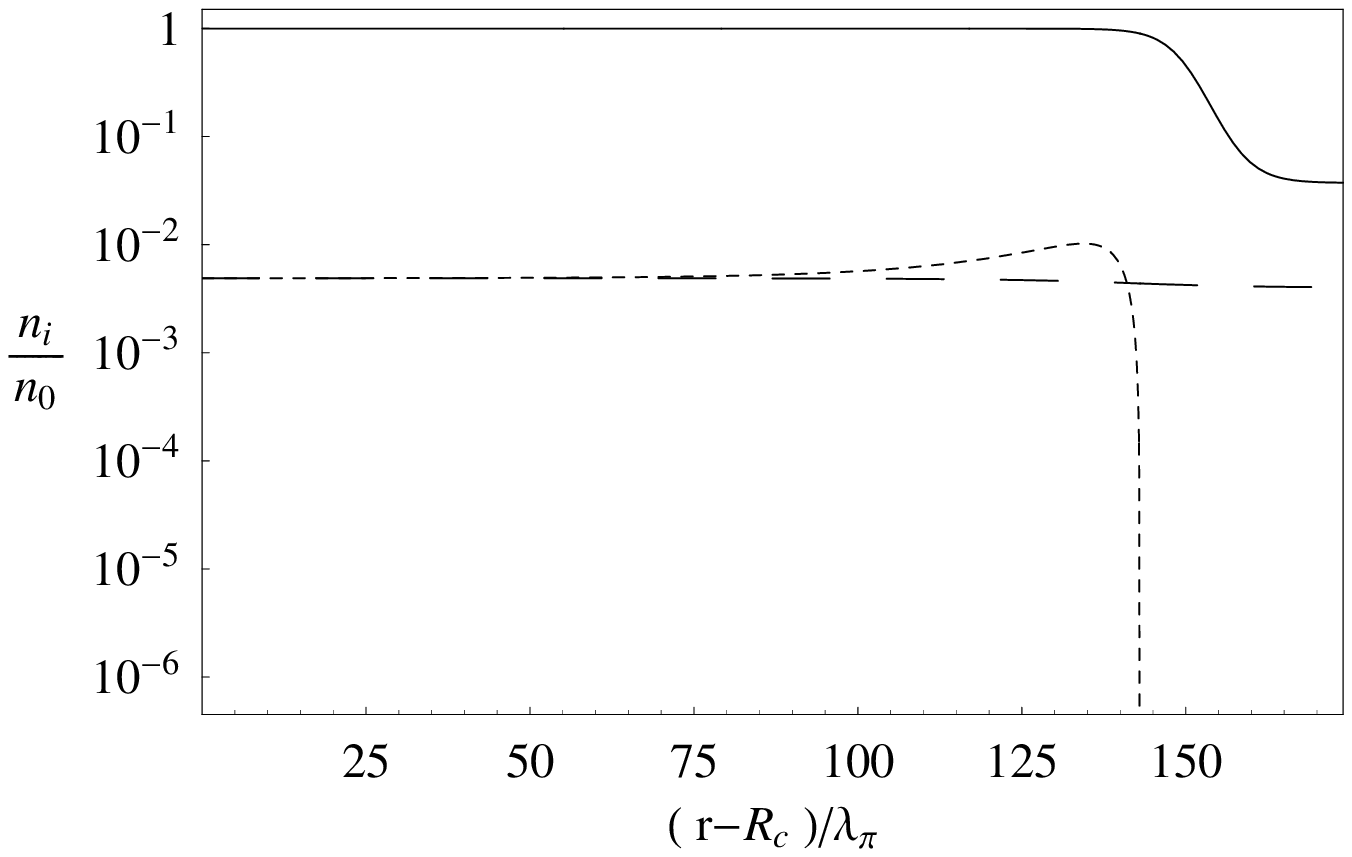}
\end{array}
$$
\caption{Left column: the surface electric field in units of the critical field. Right column: the surface particle number density of neutrons (solid), protons (short-dashed), and electrons (long-dashed) normalized to the nuclear density for selected values of $E^F_e$. First row: $E^F_e = 0.20 m_\pi$, second row $E^F_e = 0.35 m_\pi$.}\label{fig:shellE}
\end{figure}

Particularly interesting are the electrodynamical structure and the distribution of neutrons, protons, and electrons as the surface of the core is approached (see Fig.~\ref{fig:shellE}). It is interesting to compare and contrast these results with the preliminary ones obtained in the simplified model of massive nuclear density cores \cite{letter2}. The values of the electric field are quite close and are not affected by the constant proton density distribution assumed there. In the present case, the proton distribution is far from constant and increases outward as the core surface is approached.

In conclusion, for any given value of the central density an entire new family of equilibrium configurations exists. Each configuration is characterized by a strong electric field at the core-crust interface. Such an electric field extends over a thin shell of thickness $\sim 1/m_e$ and becomes largely overcritical in the limit of decreasing values of the crust mass and size (see Table \ref{table:physicalvalues} and Fig.~\ref{fig:shellE}).

These configurations endowed with overcritical electric fields are indeed stable against the quantum instability of pair creation because of the Pauli blocking of the degenerate electrons \cite{physrep}. It is expected that during the gravitational collapse phases leading to the formation of a neutron star, a large emission of electron-positron pairs will occur prior to reaching a stable ground state configuration. Similarly during the merging of two neutron stars or a neutron star and a white-dwarf leading to the formation of a black hole, an effective dyadotorus \cite{dyadotorus} will be formed leading to very strong creation of an electron-positron plasma. In both cases the basic mechanism which makes gravitational collapse depart from a pure gravitational phenomena is due to the electrodynamical process introduced in this letter. 

Finally, it is appropriate to recall that the existence of overcritical fields on macroscopic objects of $M\sim M_\odot$ and $R\sim 10$ km was first noted in the treatment of quark stars \cite{Witten, Itoh, alcock, glendenning}. In that case the relativistic Thomas-Fermi equations were also considered. However, in all of these investigations, a hybrid combination of general and special relativistic treatments was adopted, resulting in an inconsistency in the boundary conditions (see \cite{jorgePRD}). The treatment given here in this letter is the first self-consistent treatment of the general relativistic Thomas-Fermi equations, the beta equilibrium condition and the Einstein-Maxwell equations. Critical fields are indeed obtained on the surface of the neutron star core involving only neutrons, protons, and electrons, their fundamental interactions, and with no quarks present.

While we were preparing our work an extremely interesting observational problematic has emerged from the Chandra observations of Cas A CCO \cite{apj, nature}. It is with a similar steadily emitting and non-pulsating neutron star that our theoretical predictions can be tested. In particular, the existence for each central density of a new family of neutron stars with a smaller crust than the one obtained when the local neutrality condition is adopted.

Indeed, the existence of neutron stars with huge crusts, i.e., with both inner and outer crusts, is mainly a consequence of assuming no electrodynamical structure (i.e., assuming local neutrality) and of allowing electrons to have larger values of their Fermi energy $E^F_e$ (see details in \cite{jorgePRD}). It can also be demonstrated that no consistent solution of the Einstein-Maxwell equations satisfying the local $n_e=n_p$ condition exists, even as a limiting case \cite{jorgePRD}.


\begin{thebibliography}{0}
\expandafter\ifx\csname natexlab\endcsname\relax\def\natexlab#1{#1}\fi
\expandafter\ifx\csname bibnamefont\endcsname\relax
  \def\bibnamefont#1{#1}\fi
\expandafter\ifx\csname bibfnamefont\endcsname\relax
  \def\bibfnamefont#1{#1}\fi
\expandafter\ifx\csname citenamefont\endcsname\relax
  \def\citenamefont#1{#1}\fi
\expandafter\ifx\csname url\endcsname\relax
  \def\url#1{\texttt{#1}}\fi
\expandafter\ifx\csname urlprefix\endcsname\relax\def\urlprefix{URL }\fi
\providecommand{\bibinfo}[2]{#2}
\providecommand{\eprint}[2][]{\url{#2}}

\end{thebibliography}


\begin{thebibliography}{00} 

\bibitem{physrep} R.~Ruffini, S.-S.~Xue, and G.~V.~Vereshchagin, Phys.~Rep. in press (2009).

\bibitem{veresh} A.~G.~Aksenov, R.~Ruffini, and  G.~V.~Vereshchagin, Phys.~Rev.~Lett. {\bf 99}, 125003 (2007).

\bibitem{vissani1} G.~Pagliaroli, F.~Vissani, M.~L.~Costantini, and A.~Ianni, Astropart.~Phys. {\bf 31}, 163 (2009).

\bibitem{vissani2}   G.~Pagliaroli, F.~Vissani, E.~Coccia, and W.~Fulgione, Phys, Rev. Lett., {\bf 103}, 031102 (2009).

\bibitem{BBP} G.~Baym,  H.~A.~Bethe, and  C.~J.~Pethick, Nucl.~Phys.~A {\bf 175}, 225 (1971).

\bibitem{jorgePRD} Jorge A.~Rueda, R.~Ruffini, and S.-S.~Xue, To be submitted to Phys.\ Rev.\ D, (2009).

\bibitem{letter2} M.~Rotondo, Jorge A.~Rueda, R.~Ruffini, and S.-S.~Xue. To be submitted to Phys.\ Rev.\ D, (2009).

\bibitem{dyadotorus} C.~Cherubini, A.~Geralico, J.~A.~Rueda H., and R.~Ruffini, Phys.\ Rev.\ D {\bf 79}, 124002 (2009).

\bibitem{Witten} E.~Witten, Phys.\ Rev.\ D, {\bf 30}, 272 (1984).

\bibitem{Itoh} N.~Itoh, Prog.\ Theor.\ Phys.\ {\bf 44}, 291 (1970).

\bibitem{alcock} C.~Alcock , E.~Farhi, and A.~Olinto, Astrophys.\ Journal {\bf 310}, 261 (1986). 

\bibitem{glendenning} Ch.~Kettner, F.~Weber, M.~K.~Weigel, and N.~K.~Glendenning, Phys.\ Rev.\ D {\bf 51}, 1440 (1995).

\bibitem{apj} G.~G.~Pavlov and G.~J.~M.~Luna, Astrophys.\ Journal {\bf 703}, 910 (2009).

\bibitem{nature} Wynn C.~G.~Ho and Craig O.~Heinke, Nature {\bf 462}, 71 (2009).

\end{thebibliography}
\end{document}